\documentclass[twocolumn,pra,aps,nofootinbib,superscriptaddress]{revtex4-1}
\usepackage[english]{babel}
\usepackage[utf8]{inputenc}
\usepackage{url,psfrag,graphicx,color}
\usepackage{amsmath,amssymb,amsthm}
\usepackage{bm}
\usepackage{hyperref}
\usepackage{physics}
\usepackage{epsfig}
\usepackage{mathptmx}
\usepackage[english]{babel}
\usepackage{amssymb}
\usepackage{amsfonts}
\usepackage{amsmath}
\usepackage{eucal}
\usepackage{graphicx,color,placeins}
\usepackage{epsfig}
\usepackage{tikz}
\usepackage{cancel}
\usepackage{hyperref}
\usepackage[normalem]{ulem}

\def\<{\left<}
\def\>{\right>}
\def\ket|#1>{\left|#1\right>}
\def\bra<#1|{\left<#1\right|}
\def\Tr{\text{\rm Tr}}

\def\elem<#1|#2|#3>{\left<#1\right|#2\left|#3\right>}
\def\({\left(}
\def\){\right)}
\def\Z{{\mathbb Z}}
\def\C{{\mathbb C}}
\def\R{{\mathbb R}}
\def\N{{\mathbb N}}
\def\beq{\begin{equation}}
\def\eeq{\end{equation}}

\font\numbers=cmss12
\font\upright=cmu10 scaled\magstep1
\def\stroke{\vrule height8pt width0.4pt depth-0.1pt}
\def\topfleck{\vrule height8pt width0.5pt depth-5.9pt}
\def\botfleck{\vrule height2pt width0.5pt depth0.1pt}
\def\Zmath{\vcenter{\hbox{\numbers\rlap{\rlap{Z}\kern
0.8pt\topfleck}\kern 2.2pt
                   \rlap Z\kern 6pt\botfleck\kern 1pt}}}
\def\Qmath{\vcenter{\hbox{\upright\rlap{\rlap{Q}\kern
                   3.8pt\stroke}\phantom{Q}}}}
\def\Nmath{\vcenter{\hbox{\upright\rlap{I}\kern 1.7pt N}}}
\def\Cmath{\vcenter{\hbox{\upright\rlap{\rlap{C}\kern
                   3.8pt\stroke}\phantom{C}}}}
\def\Rmath{\vcenter{\hbox{\upright\rlap{I}\kern 1.7pt R}}}
\def\Hmath{\vcenter{\hbox{\upright\rlap{I}\kern 1.7pt H}}}
\def\Amath{\vcenter{\hbox{\upright\rlap{I}\kern 1.7pt A}}}
\def\Z{\ifmmode\Zmath\else$\Zmath$\fi}
\def\Q{\ifmmode\Qmath\else$\Qmath$\fi}
\def\N{\ifmmode\Nmath\else$\Nmath$\fi}
\def\C{\ifmmode\Cmath\else$\Cmath$\fi}
\def\R{\ifmmode\Rmath\else$\Rmath$\fi}

\def\max{\textrm{max}}

\def\ket|#1>{| #1 \rangle}
\def\bra<#1|{\langle #1 |}
\def\<{\langle}
\def\>{\rangle}
\def\{{\lbrace}
\def\}{\rbrace}
\def\({\left(}
\def\){\right)}
\def\[{\left[}
\def\]{\right]}

\def\AN{\mathbb{A}} 

\def\be{\begin{equation}}
\def\ee{\end{equation}}
\def\bea{\begin{eqnarray}}
\def\eea{\end{eqnarray}}
\def\Tr{{\rm Tr}}

\def\ket|#1>{| #1 \rangle}
\def\bra<#1|{\langle #1 |}
\def\<{\langle}
\def\>{\rangle}
\def\{{\lbrace}
\def\}{\rbrace}
\def\({\left(}
\def\){\right)}

\def\beq{\begin{equation}}
\def\eeq{\end{equation}}
\def\barray{\begin{eqnarray}}
\def\earray{\end{eqnarray}}

\begin{document}

\title{Entanglement as geometry and flow}  

\author{Sudipto Singha Roy}
\affiliation{Instituto de Física Teórica, UAM-CSIC, Universidad
  Aut{\'o}noma de Madrid, Cantoblanco, Madrid, Spain}

\author{Silvia N. Santalla}
\affiliation{Departemento  de Física and Grupo Interdisciplinar de Sistemas
  Complejos (GISC), Universidad Carlos III de Madrid, Spain}

\author{Javier Rodríguez-Laguna}
\affiliation{Departemento  de  Física Fundamental, UNED, Madrid, Spain}

\author{Germán Sierra}
\affiliation{Instituto de Física Teórica, UAM-CSIC, Universidad
  Aut{\'o}noma de Madrid, Cantoblanco, Madrid, Spain}

\date{\today}

\begin{abstract}
We explore the connection between the area law for entanglement and
geometry by representing the entanglement entropies corresponding to
all $2^N$ bipartitions of an $N$-party pure quantum system by means of
a (generalized) adjacency matrix. In the cases where the
representation is exact, the elements of that matrix coincide with the
mutual information between pairs of sites. In others, it provides a
very good approximation, and in all the cases it yields a natural {\em
  entanglement contour} which is similar to previous proposals.
Moreover, for one-dimensional conformal invariant systems, the
generalized adjacency matrix is given by the two-point correlator of
an {\em entanglement current} operator. We conjecture how this
entanglement current may give rise to a metric entirely built from
entanglement.
\end{abstract}

\maketitle

\section{Introduction}

Entanglement is one of the most relevant features of the quantum
world, constituting the main resource in quantum technologies
\cite{general_entang1a,general_entang1aa,general_entang1b} and
characterizing the different phases of quantum matter
\cite{general_entang2a,general_entang2aa}. In the last years, it has
been put forward that even the basic fabric of space-time might be
built upon entanglement via the holographic principle and tensor
networks \cite{M99,RT,V07,R10,S12,Cao,H17}. Indeed, this suggestive
connection stems from the {\em area law}: the entanglement entropy of
blocks of low energy states of local Hamiltonians is frequently
proportional to the measure of the boundary separating the block from
its environment \cite{area1,area2,area3,area4}, with at most
logarithmic corrections \cite{H94,V03,CC04}. Yet, there are relevant
exceptions to the area law, such as the {\em rainbow state}
\cite{VR10,rainbow_bunch1,rainbow_bunch2,rainbow_bunch3,rainbow_bunch4,rainbow_bunch5,rainbow_bunch6,rainbow_bunch7}
and the {\em Motzkin state} \cite{M1,M2,M3,M4,M5,M6,M7}. In some of
these cases, as we will show, an area law is indeed fulfilled for a
geometry that differs from the geometry defined by the local structure
of the Hamiltonian. Thus, given a quantum state, it is relevant to
ask: what is the geometry suggested by the entanglement structure?

Consider a pure state $\ket|\psi>$ of $N$ qubits.  There are $2^N$
possible subsets or blocks, $A\subset\Omega$, and we can compute the
von Neumann entropies (alternatively, Rényi entropies) for each,
$S_A=-\Tr_A(\rho_A \log \rho_A)$, where $\rho_A= \Tr_{\bar{A}}
\ket|\psi>\bra<\psi|$ and $\Tr_A$ and $\Tr_{\bar{A}}$ denote the
partial traces on subset $A$ and its complement $\bar{A}$
respectively. Let $I\in\{0,\cdots,2^N-1\}$ index the subsets through
its binary expansion, and let us compute the entanglement entropies
for all of them, $\{S_I\}_{I=0}^{2^N-1}$. The main question that we
will answer is: does this set of entropies respond to an area law for
some geometry?

This article is organized as follows. In Section \ref{sec:eam} we
define the {\em entanglement adjacency matrix} (EAM), which explains
the geometry encoded in the entanglement data, while its properties
are discussed in Sec. \ref{sec:properties}. Some exact examples are
discussed in Sec. \ref{sec:examples}. The cases, for which 
we must recourse to numerical computations, are detailed in
Sec. \ref{sec:numerical}. The EAM provides a route to define a generic
{\em entanglement contour}, as it is shown in
Sec. \ref{sec:contour}. Interestingly, from the point of view of
conformal field theory (CFT), the EAM entries can be written 
as the  two point correlator of an 
{\em entanglement current}, and many of their properties can be
readily understood. Based on the CFT insight, we ask whether an {\em
  entanglement metric} can be defined from the EAM, and a conjecture
in that direction is provided in Sec. \ref{sec:metric}. The article
concludes in Sec. \ref{sec:conclusions} with a list of our conclusions
and our proposals for further work.


\begin{figure}[t]
\includegraphics[width=8.cm]{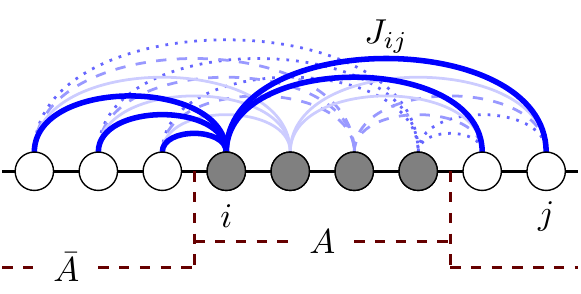}
\caption{Schematic of the entanglement entropy obtained for an
  arbitrary bipartition $(A,\, \bar A)$ by adding up the links
  connecting the sites. Here the link intensities represent the EAM
  entries, $J_{ij}$.  }
\label{fig_schematic}
\end{figure}

\section{Defining a geometry: \\ Entanglement adjacency matrix}
\label{sec:eam}

With the purpose of investigating the area law, we shall define a
geometry through an {\em adjacency matrix}, $J$, such that $J_{ij}>0$
when sites $i$ and $j$ are somehow connected or zero otherwise ($J_{ij} = J_{ji}$). The
von Neumann entropy of a subset $A\subset \Omega$ will be given by the
sum of the weights corresponding to the broken links:
\beq
S_A=\sum_{i\in A}\sum_{j\in \bar A} J_{ij},
\label{eq:SJ}
\eeq
of course, Rényi entropies of order $n$ can be employed, thus defining
$S^{(n)}_A$ and $J^{(n)}_{ij}$. Alternatively, a constant $s_0$ term
may be added to the rhs of Eq. \eqref{eq:SJ}, which may constitute a
topological entropy term \cite{note1}. If Eq. \eqref{eq:SJ} holds,
matrix $J$ will be termed the {\em entanglement adjacency matrix}
(EAM) of the state. Notice that $J_{ij}$ is the entanglement entropy
that we gain by separating node $i$ from node $j$. A schematic
representation of the above formulation is presented in
Fig. \ref{fig_schematic}. Additionally, the $J_{ij}$'s may be employed
to build a notion of distance, or an {\em entanglement metric}, along
recent proposals \cite{Cao,H17}, that we will discuss in
Sec. \ref{sec:metric}.

\begin{figure}
\includegraphics[width=8.5cm]{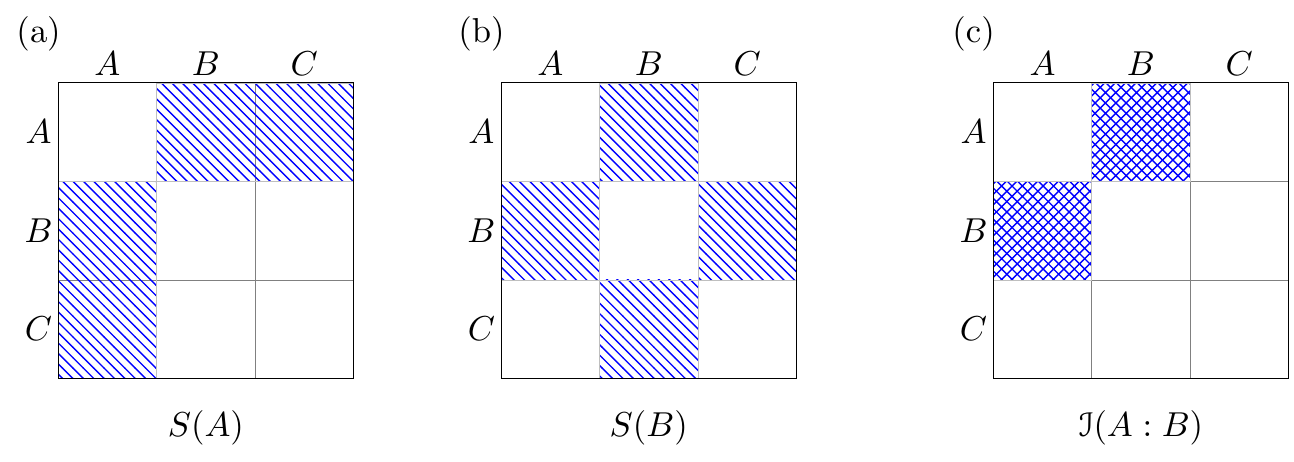}
\caption{Illustrating the structure of the entanglement adjacency
  matrix through Venn diagrams. (a) Evaluation of the entanglement
  entropy of block $A$ requires adding up all $J_{ij}$ elements
  joining $A$ to its complement ($\bar A=B\cup C$), given by the
  shaded area. (b) The same situation holds for block $B$. (c) The
  mutual information of blocks $A$ and $B$, ${\cal
    I}(A:B)=S(A)+S(B)-S(A\cup B)$ is (twice) the darker area shaded in
  this panel, given by the sum of the matrix entries $J_{ij}$ joining
  both blocks.}
\label{fig_venn}
\end{figure}

Fig. \ref{fig_venn} presents a different way to conceptualize the
entropy of any bipartition, assuming the validity of
Eq. \eqref{eq:SJ}, similar to the Venn diagrams used in classical
information theory \cite{N10}. Let the square represent the whole $J$
matrix. Then, the entropy of block $A$ or $B$ is found by counting the
shaded elements of Fig. \ref{fig_venn} (a) or (b), which are the ones
that connect the blocks with their compliments. But these quantum versions of Venn
diagrams can convey more information. The mutual information between
blocks $A$ and $B$, defined as ${\cal I}(A:B)=S(A)+S(B)-S(A\cup B)$,
which is {\em twice} the shaded area in Fig. \ref{fig_venn} (c),
i.e. the matrix elements connecting blocks $A$ and $B$.

\section{Properties of the entanglement adjacency matrix} 
\label{sec:properties}

Before considering the validity of Eq. \eqref{eq:SJ} for quantum
states of physical interest, let us list a few relevant properties of
the elements of the EAM, $J_{ab}$.

\subsection{Positivity: $J_{ab}\geq 0$.}

\proof The entanglement entropy of site $a$ can be easily obtained
(assuming $s_0=0$): $S_a=\sum_{j\neq a} J_{aj}$. Similarly, the
entropy of a block composed by sites $a$ and $b$ is given by
\beq
S_{ab}=\sum_{j\neq a,b} \(J_{aj}+J_{bj}\) = S_a+S_b-2J_{ab}.
\label{eq:s2}
\eeq
Thus, we can find the mutual information between sites $a$ and
$b$, ${\cal I}(a:b)$,
\beq
{\cal I}(a:b)\equiv S_a+S_b-S_{ab} = 2J_{ab},
\label{eq:mutual}
\eeq
thus providing a simple physical interpretation for $J_{ab}$ as the
mutual information of the pair of sites, $a$, $b$. Mutual information
must be positive, thus reinforcing our notion that $J_{ab}\geq 0$ must
hold (if $s_0=0$).

This interpretation of the entries of the entanglement adjacency
matrix as mutual information of sites yields the next corollary.

{\em Corollary: If Eq. \eqref{eq:SJ} is exact, knowledge about the
  entanglement of all single sites and all pairs of sites is enough to
  determine the entanglement of all blocks.}

\subsection{Subadditivity condition} The entanglement entropies
obtained using the elements of entanglement adjacency matrices, fulfil
certain subadditivity conditions:

{\em Given any three subsets, $A$, $B$ and $C\subset \Omega$, the
  strong subadditivity condition must hold\/\cite{N10},}
\beq
S_{AB}+S_{BC}\geq S_{ABC}+S_B.
\label{eq:strong_subadd}
\eeq
Whenever Eq. \eqref{eq:SJ} holds with $J_{ab}\geq 0$, even with
$s_0\neq 0$, this inequality will hold too.

\proof

First of all, let us consider
the blocks $A$, $B$ and $C$ contain a single site,
respectively $a$, $b$ and $c$. Then we can prove the following:

\beq
S_{ab}+S_{bc}\geq S_{abc}+S_b,
\label{eq:strong_subadd_sites}
\eeq
which is the site equivalent of Eq. \eqref{eq:strong_subadd}. In order
to prove the above relation, we can compute each of the terms:

\begin{align}
  S_b=&\sum_{j\neq b} J_{bj},\nonumber \\
  S_{ab}=&\sum_{j\neq a,b} \( J_{aj}+J_{bj}\),\nonumber\\
  S_{bc}=&\sum_{j\neq b,c} \(J_{cj}+J_{bj}\),\nonumber\\
  S_{abc}=&\sum_{j\neq a,b,c} \(J_{aj}+J_{bj}+J_{cj}\),
  \label{eq:Sabc}
\end{align}
from which we derive

\begin{align}
S_{ab}+S_{bc}=&\sum_{j\neq a,b} J_{aj} + \sum_{j\neq a,b} J_{bj} +
\sum_{j\neq b,c} J_{bj} + \sum_{j\neq b,c} J_{cj}= \nonumber\\
=&
\sum_{j\neq a,b,c}\( J_{aj}+J_{bj}+J_{bj}+J_{cj} \)+\nonumber \\  
&+ J_{ac}+J_{bc}+J_{ab}+J_{ac}=\nonumber \\
=&
S_{abc}+\sum_{j\neq a,b,c} J_{bj} + 2J_{ac}+J_{bc}+J_{ba}= \nonumber\\
=& S_{abc} + S_b+ 2J_{ac}.
\label{eq:proof_subadd}
\end{align}
Due to the positivity of $J_{ac}$, the strong subadditivity condition
is therefore proved.

\begin{figure*}
  \includegraphics[width=12cm]{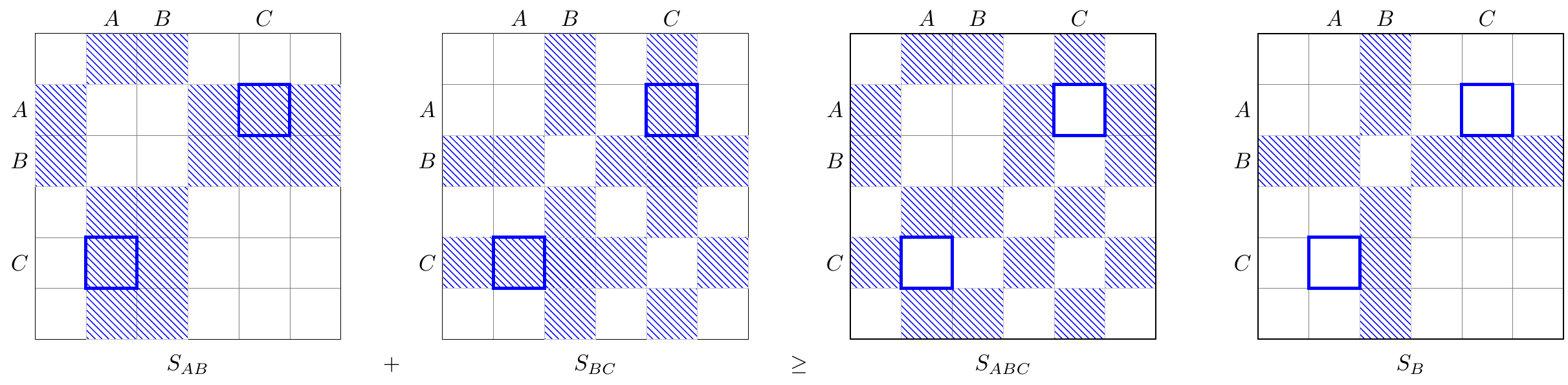}
  \includegraphics[width=12cm]{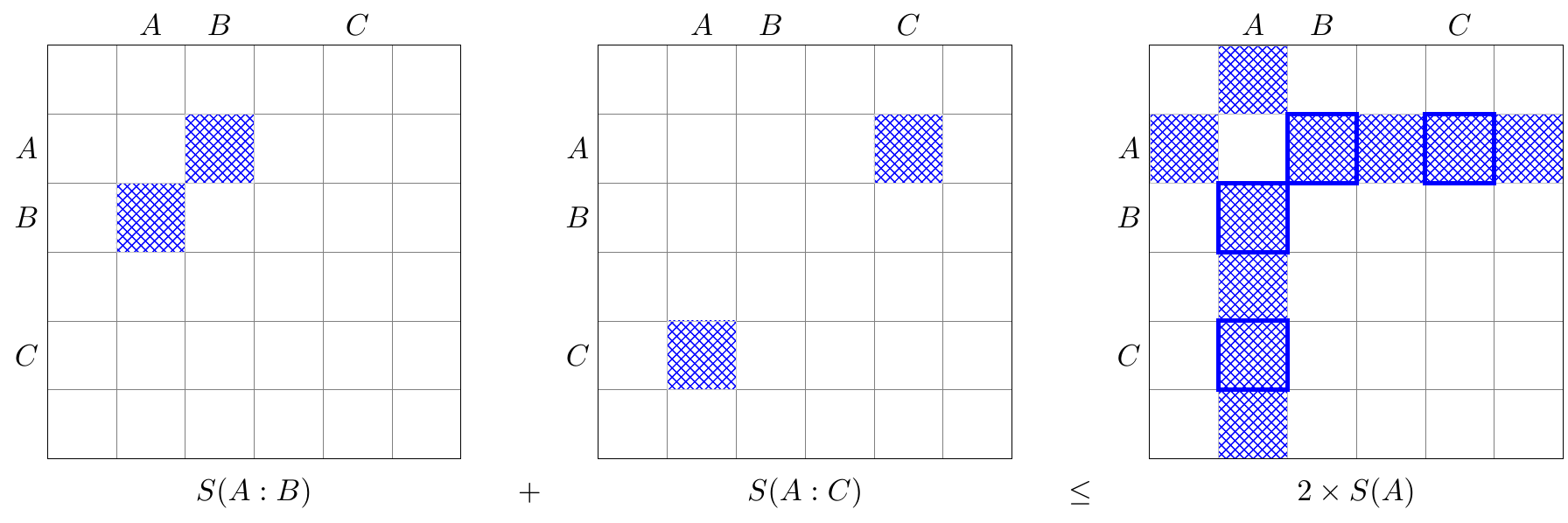}
\caption{Top: graphical illustration of the validity of
  Eq. \eqref{eq:strong_subadd}. The marked squares, filled on the lhs
  and empty on the rhs, account for the inequality. Bottom: graphical
  ilustration of the validity of Eq. \eqref{eq:IABpIAC}. The marked
  squares on the rhs are the filled squares on the lhs of the
  inequality.}
\label{sub_add_schematic}
\end{figure*}

Now let us consider general subsets $A, B$, and $C\subset \Omega$. We
can additionally prove the following important Lemma.

{\it Lemma: A partition of the nodes into blocks leads to an effective
  entanglement adjacency matrix neglecting the intra-block links and
  adding the inter-block ones.}  In other terms, if we define blocks
$\{B_k\}$, with $B_k\subset \Omega$, $B_k\cap B_l = \emptyset$ if
$k\neq l$, $\cup_k B_k=\Omega$, then we can define

\beq
J^B_{kl}\equiv \sum_{\substack{a\in B_k,\\ b\in B_l}} J_{ab},
\label{eq:JB}
\eeq
if $k\neq l$ and $J^B_{kk}=0$. Then, Eq. \eqref{eq:SJ} still holds for
partitions which do not break the blocks. The effective entanglement
adjacency matrix entries retain their physical meaning of mutual
informations. Similarly, the site entropies $S_a$, $S_{ab}$, $S_{bc}$,
$S_{abc}$ can be generalized to corresponding block entropies $S_A$,
$S_{AB}$, $S_{BC}$, $S_{ABC}$. Using this lemma, and the relation in
Eq. \eqref{eq:proof_subadd} we can prove the strong subadditivity
condition of three subsets of $\Omega$, $A$, $B$ and $C$, given in
Eq. \eqref{eq:strong_subadd}.

A graphical illustration of Eq. \eqref{eq:strong_subadd} can be seen
in Fig. \ref{sub_add_schematic}, along with an illustration of a
direct variant:

\begin{equation}
  {\cal I}(A:B)+{\cal I}(B:C)\leq 2 S(B).
  \label{eq:IABpIAC}
\end{equation}

\subsection{Recursion relation for entropies of contiguous blocks}
 If the quantum state is translationally invariant in one-dimension (1D), a
recursion relation that generalizes Eq. \eqref{eq:mutual} can be
proved for entropies of contiguous blocks, $S_l$. Since
$J_{ij}=J(l=|i-j|)$, we have

\begin{equation}
 J_l=S_l-\frac{1}{2}S_{l-1}-\frac{1}{2}S_{l+1}. 
\label{eq:JSl}
\end{equation}
The proof can be obtained by mere substitution of Eq. \eqref{eq:SJ}.

We may now ask the following question, {\it are there any real quantum
  states for which expression \eqref{eq:SJ} is exact or, at least,
  approximate?} In the forthcoming sections, we carry out analytical
and numerical analysis to answer this question in detail.


\section{Exact examples}
\label{sec:examples}

In order to obtain the entanglement adjacency matrix for any general
pure quantum many-body state, one needs to compute the entanglement of
all possible bipartitions of the state, which increases exponentially
with the size of the system. Hence, computation of the entanglement
adjacency matrix, which provides even approximate representation of
all the entropies, often requires some computational effort. We
discuss the methodology in detail in
Sec. \ref{sec:numerical}. However, in some cases in which the
entropies of the quantum state can be computed analytically, the
corresponding entanglement adjacency matrix can be obtained
straightforwardly. Below we discuss few such cases.

\subsection{Valence bond states}

Let us consider {\em valence bond states}, where qubits are paired
into maximally entangled Bell pairs,
$\ket|\Psi>=\ket|i_1,j_1>\otimes\cdots\otimes\ket|i_m,j_m>$, with $m=
N/2$ (for even N) and $|i,j\rangle= \frac{1}{\sqrt{2}} (|+-\rangle_{ij}
\pm |-+\rangle_{ij})$.  In that case, Eq. \eqref{eq:SJ} represents
exactly the entanglement entropy of every partition, as long as
$J_{ij}=\log 2$ iff $i=i_k$ and $j=j_k$ (or viceversa) for some
$k$, and zero otherwise. Interestingly, such states approximate the ground states (GS) of
strongly inhomogeneous free-fermionic Hamiltonian,

\beq
H_{\rm f-f}=-\frac{1}{2}\sum_{i,j=1}^{N} t_{ij} c^\dagger_ic_j,
\label{Free_fermion_Ham}
\eeq
where $c_i$ stands for the annihilation operator for a spinless
fermion on site $i$ and $t_{ij}$ are hopping amplitudes. Below we consider
two important members of the valence bond states, which can be derived
from the above Hamiltonian at certain limits of the hopping term
$t_{ij}$.

\begin{figure}
\includegraphics[width=8.cm]{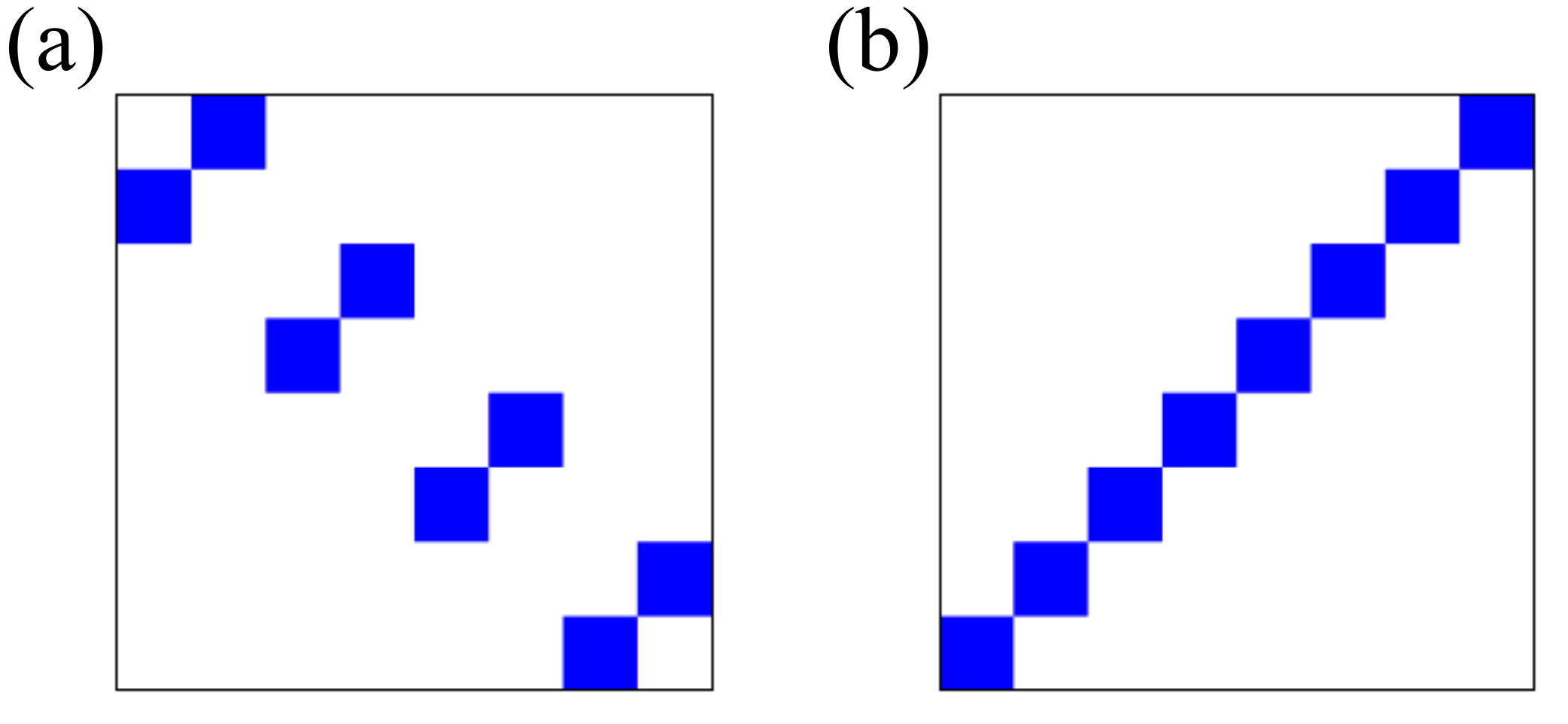}
\caption{Graphical representation of the entanglement adjacency
  matrices obtained analytically, for (a) the dimer state and (b) the
  rainbow state. Here, $N=8$.}
  \label{fig:illust}
\end{figure}

{\bf I. Dimer-model.-} If we set $t_{i,i+1}=(1+(-1)^{i+1} \delta)$, and all
other $t_{ij}=0$, for $\delta \lessapprox  1$, the GS will approximate the dimer
state, whose $J_{ij}=\log 2$ only for $i=2k-1$ and $j=2k$, for
$k=1,2,\dots\frac{N}{2}$, which is still tridiagonal, resulting in a
mere restriction of the one-dimensional adjacency matrix representing
the Hamiltonian. A graphical representation of the same is depicted in
Fig. \ref{fig:illust}(a).

{\bf II. Rainbow chain.-} Another important member of the family of
valence bond states, the {\em rainbow state}, can be derived from the
Hamiltonian in Eq. \eqref{Free_fermion_Ham}, for the following choices
of the system parameters, $t_{\frac{N}{2},\frac{N}{2}+1}=1$ and
$t_{i,i+1}=\exp(-h(|i-N/2|-1/2))$, for $h\gg 1$. In this case, the Bell
pairs are established among symmetric qubits with respect to the
center: $i_k=k$, $j_k=N+1-k$ \cite{VR10,rainbow_bunch1}. Thus, all the
entanglement entropies of the rainbow state are reproduced by
Eq. \eqref{eq:SJ} using $J_{ij}=\log 2$ iff $i+j=N+1$, and zero
otherwise. Similar to the previous example, a graphical representation
of the $J$-matrix obtained in this case is given in
Fig. \ref{fig:illust}(b). Notice that, in the rainbow case, the
entanglement adjacency matrix is not emerging as a restriction on the
adjacency matrix representing the Hamiltonian. In other words, an
observer trying to determine the geometry from observations of the
entanglement will not find the geometry of the Hamiltonian.

\subsection{Affleck-Kennedy-Lieb-Tasaki (AKLT) state}
\begin{figure}
\includegraphics[width=9cm]{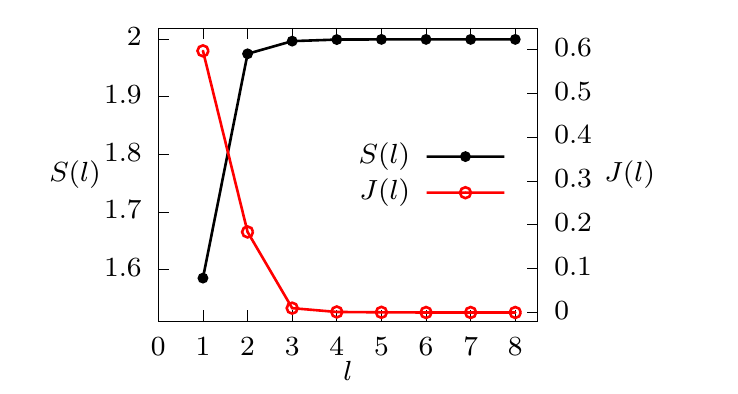}
\caption{Plot of $S(l)$ given in Eq.(\ref{AKLT_entropy}), along with
  the $J(l)$ obtained using Eq.(\ref{AKLT_entropy}) into
  \eqref{eq:JSl}. Both results are expressed taking logs in base 2.}
  \label{fig:AKLT}
\end{figure}

  Let us now consider a translational invariant quantum state,  the Affleck-Kennedy-Lieb-Tasaki (AKLT) state on a spin-1  chain \cite{AKLT} and derive the exact expression of the
$J(l=|i-j|)$ as follows. The expression of entanglement entropy for a
block of $l$ consecutive sites obtained from the periodic AKLT state
with $N\gg 1$ 
is given by \cite{periodic_AKLT_ent,AKLT_red1,AKLT_red2}

\begin{equation}
S_l= \log4-\frac{3}{4}(1-p^l)\log(1-p^l)-\frac{1}{4}(1+3p^l)\log(1+3p^l),
\label{AKLT_entropy}
\end{equation}
where $p=-\frac{1}{3}$. For $l\gg 1$,  $S_l$  approaches asymptotically 
the value $2 \log 2$ that corresponds to cutting the  valence bonds 
that connect the block to the rest of the system (see Fig. \ref{fig:AKLT}). 
$J(l)$ can be found plugging (\ref{AKLT_entropy}) into 
the recursion relation for translationally
invariant states,  \eqref{eq:JSl}.
The  result is plotted in Fig. \ref{fig:AKLT}.  Notice the fast decreases of $J(l)$ with
$l$, that  can be obtained expanding  the entropies in
Eq. \eqref{AKLT_entropy} as

\begin{align} 
  S_l \approx &\log4-\frac{3}{4}(1-p^l)
  \Big[-p^l-\frac{1}{2}(p^l)^2\Big] \nonumber\\
  &-\frac{1}{4}(1+3p^l)\Big[3p^l-\frac{1}{2}(3p^l)^2\Big].\nonumber\\
  \approx &\log4-\frac{3}{2}p^{2l}. 
\end{align}
Replacing  this into  Eq. \eqref{eq:JSl}, and only keeping the terms upto
$p^{2l}$, we get

\begin{align}
  J_l
  =&\log4-\frac{3}{2}p^{2l}-\frac{1}{2}\log 4+\frac{1}{2}\(\frac{3}{2}p^2p^{2l}\) \nonumber\\
  &-\frac{1}{2}\log4+\frac{1}{2}\(\frac{3}{2}p^{-2} p^{2l}\),
\end{align}
which after simplification yields 

\begin{equation}
 J_l \propto p^{2l}.
\label{eq:J_AKLT}
\end{equation}
Therefore, for AKLT state, $J_l$ decays faster than that of
spin-correlation function, $\langle \vec{S}_0 \vec{S}_l\rangle = 4
p^l$ \; \cite{AKLT}.

\subsection{GHZ-state}

A different case is provided by the Greenberger-Horne-Zeilinger (GHZ)
state, $\ket|\psi>_{\rm GHZ}=\frac{1}{\sqrt{2}}(|0\rangle^{\otimes
  N}+|1\rangle^{\otimes N})$. In this case, the entanglement entropy
of all partitions is equal to $\log 2$. It can be proved that
Eq. \eqref{eq:SJ} can only represent this situation making $J_{ij}=0$
for all $i,j$ and $s_0=\log 2$. This amounts to the fact that the GHZ
state does not have a geometrical interpretation in this framework.


\section{Numerical computation of the entanglement adjacency matrix}
\label{sec:numerical}

Eq. \eqref{eq:SJ} attempts to represent $2^N$ entanglement entropy
values using only $N_p=N(N-1)/2$ parameters (neglecting $s_0$). The
relation between parameters and entropies is linear, expressed through
\beq
\sum_{(ij)}{\cal A}_{I,(ij)} J_{ij} = S_I,
\label{eq:defA}
\eeq
where $I=(x_1\cdots x_N)$ denotes the binary expansion for the index
of each block, i.e. $x_k=1$ if site $k$ belongs to block $I$ (and zero
otherwise), and ${\cal A}$ is a $2^N \times N_p$ matrix with entries
given by ${\cal A}_{(x_1\cdots x_N),(ij)}=1$ if $(x_i,x_j) =(0,1)$ or
(1,0), and zero otherwise. Eq. \eqref{eq:defA} is a strongly
overdetermined linear system which will be, in general,
incompatible. Yet, it is possible to find an approximate solution in
the least-squares sense, using the so-called normal equations

\beq
\sum_{(i'j')} \( {\cal A}^\dagger {\cal A} \)_{(ij),(i'j')} J_{i'j'}=
  \sum_I {\cal A}_{I,(ij)} S_I,
\label{eq:normaleqs}
\eeq
So, element $kl$ of matrix $\mathcal{A}^{T} \mathcal{A}$ provides the
number of blocks which break both index $k$ and index $l$. This number
is independent of $k$ and $l$ as long as $k\neq l$. For a system with
$N$ sites, the number of blocks which break a given coupling is always
the same, $2^{N-1}$. The number of blocks which break two given
couplings is also the same: $2^{N-2}$. Thus, matrix $\mathcal{A}^T
\mathcal{A}$ is given by

\beq
\mathcal{A}^T\mathcal{A}=2^{N-2}
\begin{bmatrix}
2 & 1 &  1&\dots& 1\\
1&2&1 & \dots &1\\
1&1&2 & \dots &1\\
\vdots & \vdots&\vdots&\ddots\\
1&1&\dots &\dots&2\\
\end{bmatrix}.
\eeq

Eq. \eqref{eq:normaleqs} is a linear system of $N_p$ equations for
$N_p$ unknowns with a unique solution, but the computational cost is
still exponential because it requires the evaluation of $2^N$
entropies. Nevertheless, an approximate solution can be found using a
random sample of the total set of entropies.

We quantify the relative error made in the optimization process
described above, as follows. Let $\hat S_I$ be the estimate obtained
through Eq. \eqref{eq:normaleqs}. The error will be defined as
\beq
   {\cal E}={1\over 2^N} \sum_{I=0}^{2^N-1} \left| S_I- \hat S_I \right|.
\label{eq:error}
\eeq
We will use this formula to estimate the error made in the computation
of the entanglement adjacency matrix for certain physical models we
consider below.

\subsection{Free-fermionic model}

Let us consider free-fermionic systems, as described in
Eq. \eqref{Free_fermion_Ham}. The GS of \eqref{Free_fermion_Ham} is a
Slater determinant built from the lowest energy eigenstates of the
hopping matrix $t_{ij}$. Let $U_{k,i}$ be the matrix containing such
eigenstates as columns. Then,

\beq
C_{ij}=\<c^\dagger_i c_j\>=\sum_{k\in K} \bar U_{k,i} U_{k,j},
\label{eq:corr}
\eeq
where $K$ denotes the set of occupied orbitals, which we will assume
to be the half with negative energies. The von Neumann entropy for a
block $A$ is found from the eigenvalues of the restriction of $C_{ij}$
to that block, $\nu_p\in [0,1]$ \cite{P03}

\beq
S_A=\sum_{p=1}^{|A|}  H(\nu_p), \quad 
H(x) = -  \left[ x \log x + (1- x)\log (1-x) \right]. 
\label{eq:fermion_S}
\eeq

\begin{figure}[h]
  \includegraphics[width=7cm]{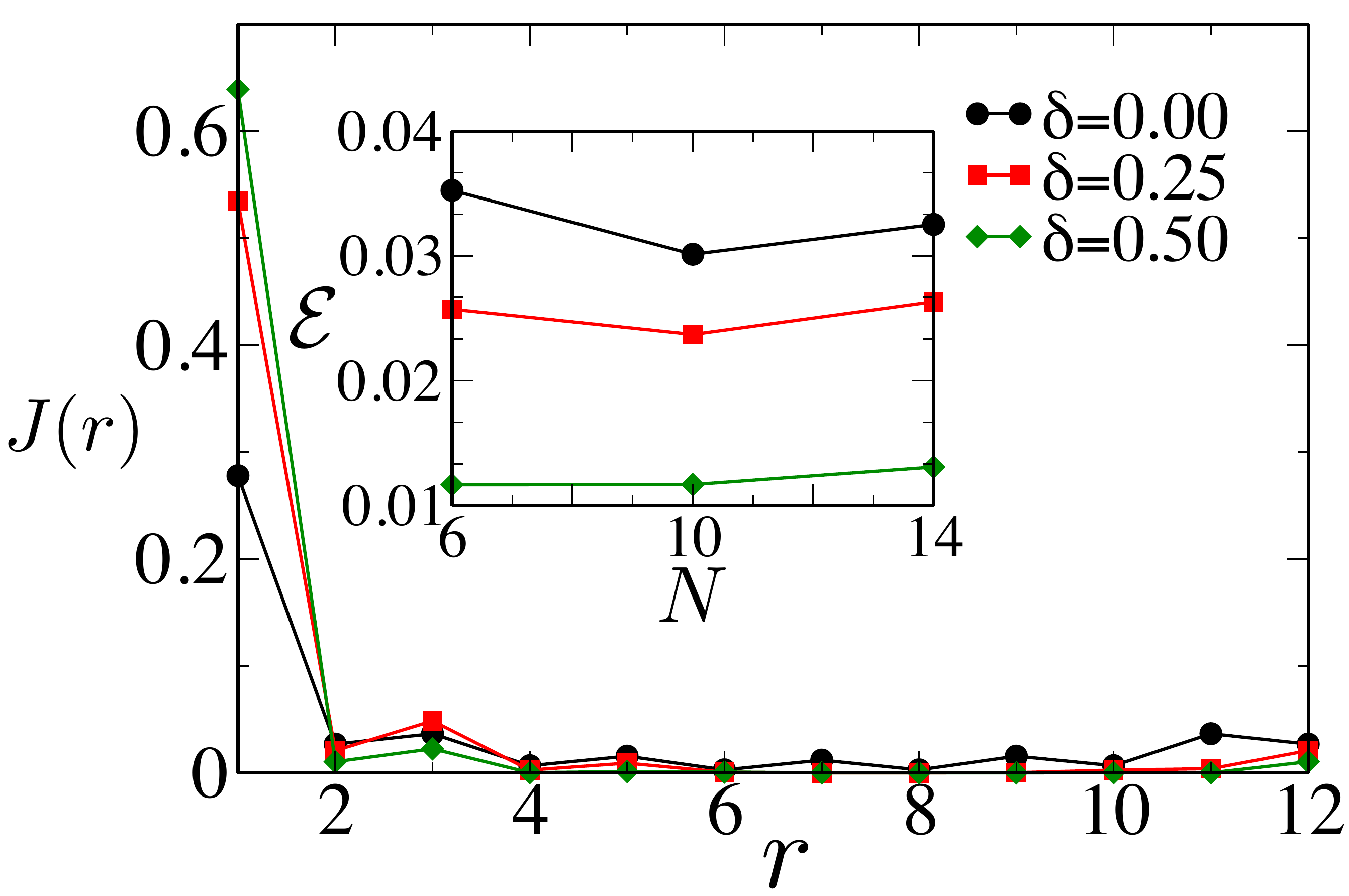}
  \caption{ We consider the nearest-neighbor dimerized Hamiltonian
    with periodic boundary condition and $N=14$. Plots of
    $J(r)=J_{2,2+r}$ with the distance $r$ are obtained for different
    $\delta$ values. Additionally, in the inset, we plot the scaling
    of the error function $\mathcal{E}$ with the system size $N$, for
    the same choices of the parameters.}
  \label{label:fig2}
\end{figure}

Once the entropy values are computed, the optimal entanglement
adjacency matrix can be obtained by solving the set of linear
Eqs. \eqref{eq:normaleqs}. In Fig. \ref{label:fig2} we plot the
$J(r)=J_{2,2+r}$ with the distance $r$, for the dimerized Hamiltonian,
which can be derived from the free-fermionic model expressed in
Eq. \eqref{Free_fermion_Ham}, for $t_{i,i+1}=1+(-1)^{i}\delta$. From the
figure, we can see that starting from a high value, $J(r)$ decreases
with $r$. Moreover, $J(r)$ presents slowly decaying parity
oscillations. A true short-ranged behavior emerges as the GS
configuration tends towards the dimer configuration for higher values
of $\delta$. Subsequently, we present a scaling of the error
($\mathcal{E}$) estimated in all these cases with the system size
$N$. The maximum error turns out to be $< 4\times10^{-2}$.  Though $\mathcal{E}$ quantifies the average error introduced in the computation of entropy of any bipartition, a more fine-grained analysis of the error, in this case, reveals the fact that the amount of error is relatively lower in case of bipartitions with less number of boundaries (see  Appendix \ref{Appendix_A} for a detailed discussion ). 

\subsection{Interacting model: XXZ Hamiltonian}

\begin{figure}[h]
  \includegraphics[width=7cm]{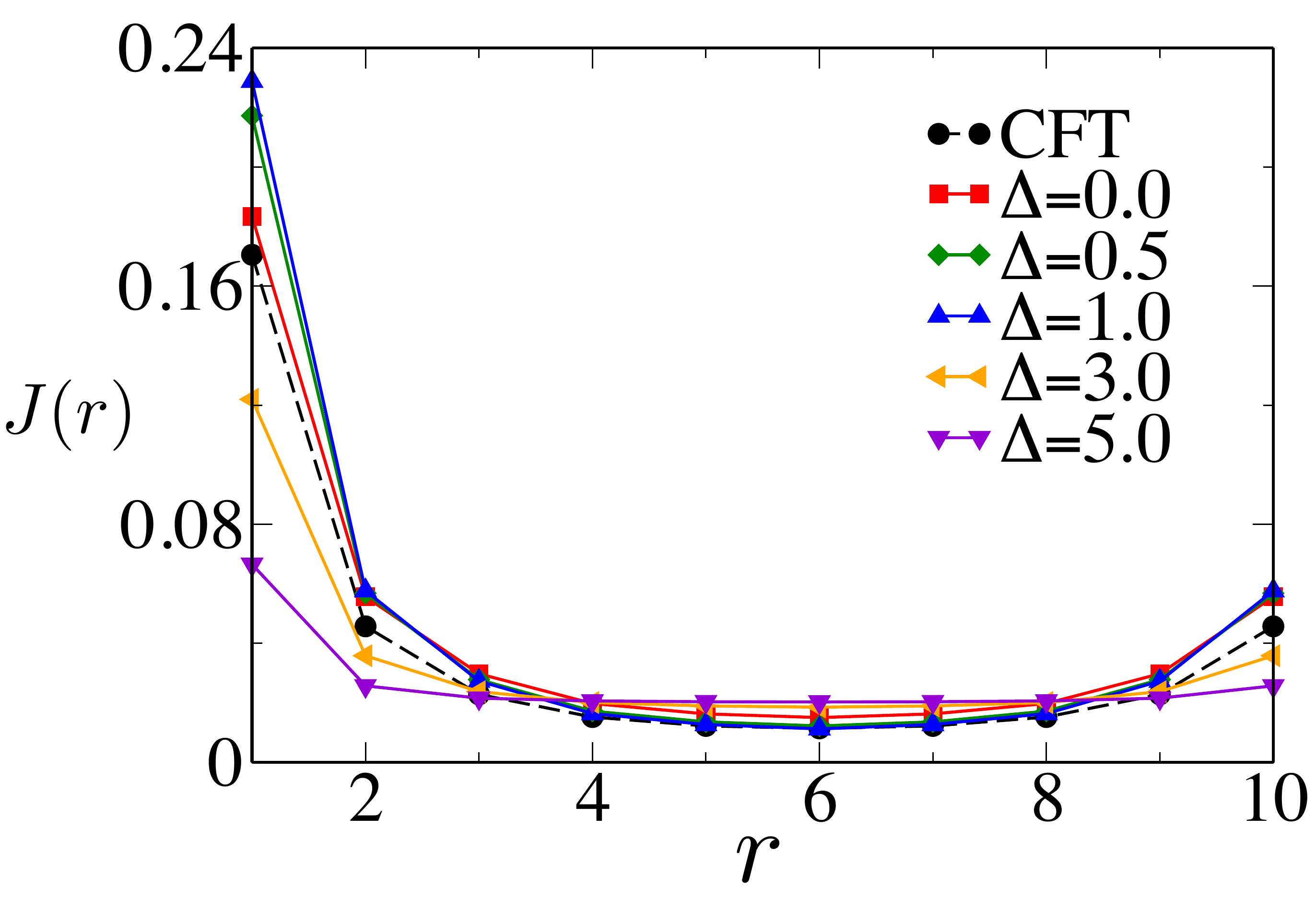}
  \caption{ We obtain the same profiles for the $XXZ$ model expressed
    in Eq. (\ref{eqn:xxz}), with $N=12$ and for different values of
    the anisotropy parameter $\Delta$. The broken black line
    corresponds to the analytical expression $J(r) = \frac{1}{6}
    (\frac{\pi}{N})^2 \frac{1}{ \sin^2(\pi r/N)}$, obtained for the
    conformal case (see Table I in Appendix \ref{Appendix_C})}
\label{label:fig3}
\end{figure}

We next discuss the optimal entanglement adjacency matrix obtained for
an interacting model, the $XXZ$ model in 1D with periodic boundaries,
expressed as

\beq
H_{XXZ}=\sum_i^N \left[S_i^xS_{i+1}^x+S_i^yS_{i+1}^y+
  \Delta S_i^zS_{i+1}^z \right],
\label{eqn:xxz}
\eeq
where $S_i^k$ ($k\in {x,y,z}$) are the spin-1/2 operators at site $i$,
and $\Delta$ denotes the anisotropy constant. We carry out the same
analysis as before and obtain the entanglement adjacency matrix by
solving the set of linear Eqs. \eqref{eq:normaleqs}. However, in this
case, the entropy values cannot be obtained from the correlation
matrix. Rather, we need to perform an exact diagonalization and
obtain the reduced density matrices directly from the GS itself. The
behavior of $J(r)=J_{2,2+r}$ with the distance $r$ is depicted in
Fig. \ref{label:fig3}, for the following values of the anisotropy
parameter $\Delta=0.0, 0.5, 1.0$ (critical cases), and $3.0,5.0$
(non-critical cases in the gapped antiferromagnetic phase).

In general, spin and the fermion representations lead to different
reduced density matrices for non-consecutive blocks \cite{Peschel}.
In our case, in order to obtain the optimized geometry, we must
consider both consecutive and non-consecutive blocks, a clear
difference emerges in the behavior of $J(r)$ obtained in this case. We
observe that unlike the fermionic case, the profile of $J(r)$ obtained
using the spin representation does not exhibit the usual parity
oscillations. In this case, the maximum error ($\epsilon$) computed
using the formula in Eq. (\ref{eq:error}), turns out to be $\sim
6\times 10^{-2}$. Moreover, here the random sampling of only $5\times 10^2$ entropies provides close agreement with the above results (see Fig. \ref{fig:random_sampling} in Appendix \ref{Appendix_A}). Hence, the random sampling method can ensure the scalability of the formalism and unveil important features at larger system sizes, when exploring all possible bipartitions becomes impracticable.


\section{Entanglement contour}
\label{sec:contour}

In the subsequent part of our analysis we use the entanglement
adjacency matrix to refine the entanglement structure contained in the
block entropies, making use of the {\em entanglement contour} function
\cite{V14}. The entanglement contour for a given block $A$, introduced
by Chen and Vidal \cite{V14} and analysed in several other works
\cite{B04,FR15,T17,Q18,Tcon,R19,rainbow_bunch4}, is a positive
partition of the entanglement entropy associated to the block sites,
i.e. a function $s_A(i)$ with $i \in A$, such that
\beq
S_A = 
\sum_{i \in A} s_A(i), \qquad s_A(i) \geq 0 \, . 
\label{eq:def_contour}
\eeq
Interestingly, the entanglement adjacency matrix provides a natural
entanglement contour, using expression \eqref{eq:SJ},
\beq
\quad s_A(i)\equiv \sum_{j\in\bar A} J_{ij}  \, . 
\label{eq:contour_J} 
\eeq
Furthermore, using the properties of $J_{ij}$ described in
Sec. \ref{sec:properties}, it can be shown that our proposal for the
entanglement contour function satisfies all the constraints listed in
Ref. \cite{V14}. Below we list-down a few of those relevant
properties.

\begin{enumerate}
\item {\it Positivity}: $s_A(i)\geq 0$.

Eq. 
\eqref{eq:mutual}
guarantees the positivity of the elements of the
entanglement adjacency matrix, $J_{ij}$, which implies that the
contour function $s_A(i)$ defined in Eq. \eqref{eq:contour_J} must be
positive.

\item {\it Normalization}: $\sum_{i\in A} s_A(i)=S_A$.

From Eq. \eqref{eq:SJ} and Eq. \eqref{eq:contour_J}, we obtain the
entropy of the subset $A$ is given by $S_A=\sum_{i\in A}\sum_{j\in
  \bar A} J_{ij}=\sum_{i\in A} s_A(i)$. Hence, the normalization
condition of the contour function is guaranteed.

\item {\it Symmetry}: If $\mathcal{F}$ is a symmetry of $\rho_A$,
  ($\mathcal{F} \rho_A \mathcal{F}^{\dagger} = \rho_A$)
that exchanges the sites $i$ and $j$, then $s_A(i) = s_A(j)$. 

This statement applies to space reflections, translations, and
rotations. In our case, the symmetry must apply to the whole
wavefunction, $\ket|\psi>_{AB}$, not to the reduced density matrix of
a single block, since the $J_{ij}$'s is a property of the entire set of partitions. A symmetry of the wavefunction will be reflected in
a symmetry of the set of entanglement entropies of the different
blocks. The solution to the normal equations,
Eq. \eqref{eq:normaleqs}, is unique and must reflect the symmetries of
the wavefunction.

\end{enumerate}

Proof of  properties, viz. invariance of entanglement
contour under local unitaries and its upper and lower bounds are given
in Appendix \ref{Appendix_B}.

\begin{figure}[h!]
  \includegraphics[width=7cm]{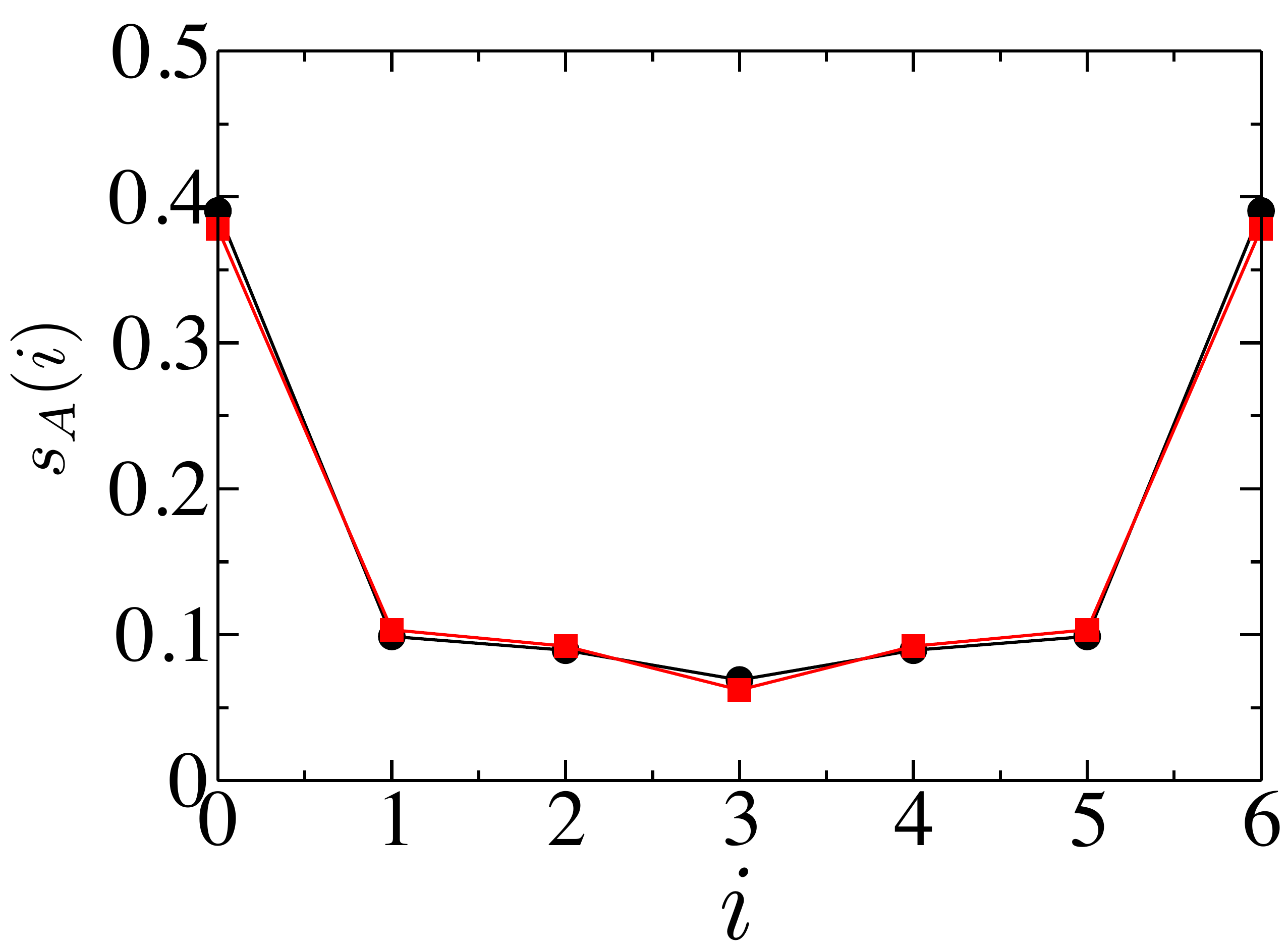}
  \caption{Comparison of the contour functions for the entanglement
    entropy $s_A(i)$, obtained for the half-chain of a clean
    free-fermionic system ($t_{ij}=1, |i-j|=1$ in
    Eq. \eqref{Free_fermion_Ham}), using Eq. \eqref{eq:contour_J} (red
    squares) and Eq. \eqref{eq:vidal_contour} (black circles). Here,
    we consider $N=14$.}
  \label{label:fig4}
\end{figure}

For free-fermion models, a proposal for the contour is \cite{V14} 
\beq
s_A(i)= \sum_{p=1}^{|A|} |\Phi_{p,i}^{(A)}|^2  \;  H(\nu_p) \, , 
\label{eq:vidal_contour}
\eeq
where $\Phi_{p,i}^{(A)}$ is the eigenvector, with eigenvalue $\nu_p$,
of the correlation matrix \eqref{eq:corr} restricted to the block $A$.
In this regard, here, we stress the fact that unlike the
  above formulation, our approach aims to provide an entanglement
  contour function by considering contributions of all bipartitions
  and not just the ones consisting of simply connected
  intervals. Moreover, the formalism can be applied to any general
  quantum system, including the interacting models.
Fig. \ref{label:fig4} shows that the contours \eqref{eq:contour_J} and
\eqref{eq:vidal_contour} for a free-fermion model are very similar \cite{note2}.

\section{Entanglement current}
\label{sec:current}

The entanglement entropy of the GS  of a conformal field
theory (CFT) for an interval $A = (u, v)$ embedded in the infinite
line is given by \cite{H94,V03,CC04}
\beq
S_A = \frac{c}{3} \log \frac{v - u}{\epsilon} \, , 
\label{cft1_text}
\eeq
where $c$ is the central charge and $\epsilon > 0$ a short distance 
cut-off. Eq. (\ref{cft1_text}) can be obtained from a continuous
version of (\ref{eq:SJ})
\beq
S_A =   \int_{A_\epsilon} dx \int_{B} dy \; J(x,y)  \, , 
\label{cft2_text}
\eeq
with  $A_\epsilon =( u+ \epsilon, v - \epsilon)$ and $B = (- \infty,
u) \cup (v, \infty)$, by  choosing
\beq
 J(x,y)  = \frac{c/6}{(x-y)^2}  \, . 
\label{cft3_text}
\eeq
Note that the above choice of the entanglement adjacency matrix is also emanating from the recursion relation given in Eq. (10), when the continuum limit is considered. In particular, if we expand the entropies obtained for the contiguous blocks, $S_l$, we get $J_l=-\frac{1}{2}\frac{d^2 S_l}{dl^ 2}$. 
Eq. (\ref{cft3_text})  indicates that $J(x,y)$ is the two point correlator, on
the plane of the spatial component of a current operator ${\bf J}$, whose integration
along segments, as in Eq. (\ref{cft2_text}), is invariant under
reparametrizations. $J(x,y) dx dy$ represents the amount of
entanglement between the intervals $(x, x+dx)$ and $(y, y+dy)$. This
interpretation of ${\bf J}$ holds in more general situations. Indeed,
using the construction by Cardy and Tonni for entanglement
Hamiltonians in CFT \cite{CT16}, we can show that
Eq. (\ref{cft2_text}) reproduces the values of $S_A$, for the
space-time geometries $\Sigma$, that are conformally equivalent to an
annulus. In these cases $J(x,y)$ is given by the two point correlator
(see Appendix \ref{Appendix_C}).

\beq
 J(x,y)  = \langle {\bf J}(x) \,  {\bf J}(y) \rangle_\Sigma  \, . 
\label{cft5_text}
\eeq 
The origin of the entanglement current ${\bf J}$ can be traced back 
to  the way  the  $n^{\rm th}$ R\'enyi  entropy, 
$S^{(n)}_A = \frac{1}{1 -n} \log {\rm tr}_A  \; \rho^n_A$, is computed 
using  twist fields \cite{CC04,CC09,CCD08}. For  the interval $A= (u,v)$
in the infinite line one has 
\beq
{\rm tr}_A  \; \rho^n_A = \langle   \bar{{\cal T}}_n(u)  {\cal T}_n(v) \rangle  =
\frac{c_n}{ ((v - u)/\epsilon))^{ 2 \Delta_n} }  \, , 
\label{twist}
\eeq
where ${\cal T}_n(x)$, and its conjugate $\bar{\cal T}_n(x)$,  are
twist fields with the same  scaling dimension
$\Delta_n = \frac{c}{12} ( n - 1/n)$,  and $c_n$ is a 
non universal constant whose value at  $n=1$ is  $c_1=1$ 
to guarantee  the normalization condition ${\rm tr}_A \; \rho_A =1$. 
The von Neumann entropy $S_A$, given in (\ref{cft1_text}),  
can be  derived  as 
$\lim_{n \rightarrow 1} S^{(n)}_A$ using (\ref{twist}). There is also 
an additive  constant constant $c'_1$, not included in (\ref{cft1_text}), 
that  comes  from the derivative of $c_n$ at $n=1$.  

The relation between  the entanglement current
and the twist fields is based on   Eq.(\ref{twist}). Taking derivatives
respect to the coordinates $u$ and $v$ one finds, in the limit $n \rightarrow 1$
\beq
\lim_{n \rightarrow 1} \frac{1}{ 2 (1-n)}  \langle   \partial_u  \bar{{\cal T}}_n(u)  \partial_v  {\cal T}_n(v) \rangle  =
\frac{c/6}{ (v - u)^{ 2} }  \, , 
\label{twist2}
\eeq
that compared with (\ref{cft3_text}) yields  the formal identification  
\beq
{\bf J}(x) = \lim_{n \rightarrow 1}  \frac{1}{ \sqrt{ 2 (1-n)}} \partial_x  {\cal T}_n(x)  \, , 
\label{twist3}
\eeq
and the same expression with $\bar{{\cal T}}_n(x)$.  It is important to observe that exchanging the limit $n \rightarrow 1$ and  the $u$ and $v$ derivatives in Eq. (\ref{twist2}) is not well defined since $\lim_{n \rightarrow 1} \langle  \bar{{\cal T}}_n(u) {\cal T}_n(v)   \rangle/(1-n) = \infty$. There is no a priori reason why these two operations should commute. The order of operations we have chosen in Eq. (\ref{twist2})  is perfectly consistent and allow us to identify the entanglement current  (\ref{twist3}). This derivation is certainly different from the standard one where the entanglement entropy is found by taking the logarithm of Eq. (\ref{twist}) and the limit $n \rightarrow 1$.  Notice that the twist
fields become the identity in the limit $n \rightarrow 1$, so their derivative
are fields of dimension 1.

Applying
(\ref{cft2_text}) and (\ref{cft3_text}) to disconnected intervals
on a line gives the formula derived for $S_A$ in \cite{CC04} (see 
Appendix \ref{Appendix_C}), 
 but misses a term that depends on the
harmonic ratio of the entangling points
\cite{CG08,F09,cct-09,cct-11,A10,coser-many,C18}. 
The current operator can be written as $J_\mu =
\partial_\mu \phi$, where $\phi$ is a massless boson, that implies its
conservation, i.e. $\partial_\mu J_\mu =0$.  If the scalar field
$\phi$ has a mass $m$, one can derive from (\ref{cft2_text}) the
entropy $S_A \simeq \frac{c}{3} \log (\xi/\epsilon)$, that corresponds
to a massive field theory in the scaling limit with correlation length
$\xi = 1/m$ \cite{CC04}.  Eqs. (\ref{cft2_text}) and (\ref{cft5_text})
can be generalized to models in $D$ spatial dimensions recovering the
area law for $S_A$.

Eq. (\ref{eq:mutual}) and Eq. (\ref{cft3_text}) seem to suggest that
the mutual information ${\cal I}(x:y)$ in CFT has a universal scaling
behaviour $|x- y|^{-2}$.  This property holds for the free-fermion
model studied above but not in general. A recent example was studied
in ref.\cite{H18}, when ${\cal I}(x:y)$ is computed for certain
spin-Hamiltonians. In this case, it is reported that ${\cal I}(x:y)
\propto |x-y|^{-\eta}$ with $\eta = 1/2$ for the critical Ising model,
and $\eta=1$ for the XX spin model \cite{H18}. The latter results does
not contradict Eq. (\ref{cft3_text}), since the spin representation
may exhibit a different decay rate of bipartite entanglement
\cite{Peschel}.


\section{Towards an entanglement metric}
\label{sec:metric}

Let us consider the intriguing possibility that the $J_{ij}$ might be
employed to build a metric entirely based on entanglement properties,
as other authors have recently explored \cite{Cao,H17}. The basic
assumption is that highly entangled sites are, somehow, {\em
  nearby}, while disentangled sites are further away. We suggest that
$J_{ij}$ determines univocally a {\em single-step} distance between
sites $i$ and $j$, through an unknown function

\begin{equation}
  d_{ij}=\Phi(J_{ij}),
  \label{eq:dJ}
\end{equation}
which is monotonously decreasing and fulfills $\Phi(0)\to\infty$ and
$\Phi(J_\max)=\ell_0$, where $J_\max$ is the maximal possible value
of  $J$ (for qubits, $J_\max=\log 2$) and $\ell_0$ is the minimal
length scale. A reasonable choice, following our CFT discussion, is

\begin{equation}
  \Phi(J)=\ell_0 (J/J_\max)^{-1/2}.
  \label{eq:phiJ}
\end{equation}

In analogy to discrete metric problems, such as
first-passage percolation (FPP) \cite{Kesten}, the actual {\em
  distance} between sites $i$ and $j$ is found by obtaining the
discrete geodesic, which we proceed to define. Let $\Gamma$ be any
path along the system, $\Gamma=\{i_1,\cdots,i_M\}$. Then,

\begin{equation}
  D_{ij}=\min_\Gamma \sum_{k=1}^{M(\Gamma)-1} d(i_k,i_{k+1}),
  \label{eq:geodesic}
\end{equation}
where $M(\Gamma)$ is the number of steps of the path $\Gamma$. It is
easy to prove that $D_{ij}$ fulfills the distance axioms, including
the triangle inequality. Notice that the geodesic can take as many
steps as required.

Let us consider a translationally invariant 1D system following the
area-law: entanglement across any  cut is bounded by a constant, with $J_{ij}=J_0\exp(-|i-j|/\xi)$, such as the AKLT state,
see Eq. \eqref{eq:J_AKLT}. Then, for sites $i<j$
such that $|i-j|\gg\xi$, we see that the geodesic is the straight path
$\Gamma=\{i,i+1,\cdots, j\}$, with a distance $D_{ij}\propto
|i-j|$. On the other hand, for conformal systems, the straight path
competes with the single-step path, which is also of order
$D_{ij}=J_{ij}^{-1/2}\sim |i-j|$, see Eq. \eqref{cft3_text}.


\section{Conclusions and further work}
\label{sec:conclusions}

To summarize, in this work, we introduced a framework to unveil the
geometry suggested by the entanglement structure of any quantum
many-body state. The optimal geometry is characterized by the elements
of a generalized adjacency matrix, which is obtained by exploring the
entanglement entropies computed for all possible bipartitions of the
many-body state.  We noted that in some cases, the optimal geometry
turns out to be completely different from that suggested by the parent
Hamiltonian of the model.  We later showed how the optimized
geometries can provide a natural route to compute the entanglement
contour, introduced for non-interacting models.  Finally, we showed
that for a conformal invariant system, the elements of the generalized
adjacency matrices can be related to the two-point correlator of an
entanglement current operator. This field theory realization leads to
think of entanglement as a flow among the parts of the system, in
analogy to the flow of energy that is characterized by the stress
tensor. Both entanglement and energy are, after all, fundamental
resources of a physical theory \cite{Cirac05}. It will also be
interesting to analyze the relation of our approach to the geometry
proposed for tensor networks \cite{H17}, to holography in static and
dynamic scenarios \cite{E10,R19,FH17} and also to higher-dimension
physical system. As a possible application of the formalism to other
physical models, we plan to explore quantum disordered and quantum
quenched systems in our future works.

\acknowledgements

We would like to thank P. Calabrese, J. I. Cirac, J. I. Latorre, E.
L\'opez, L. Tagliacozzo, E. Tonni, G. Vidal and H. Q. Zhou, Q. Q. Shi and S.Y. Cho for
conversations.  We acknowledge financial support from the grants
FIS2015-69167-C2-1-P, FIS2015-66020-C2-1-P, PGC2018-095862-B-C21,
PGC2018-094763-B-I00, QUITEMAD+ S2013/ICE-2801, SEV-2016-0597 of the
``Centro de Excelencia Severo Ochoa" Programme and the CSIC Research
Platform on Quantum Technologies PTI-001.


\appendix
\begin{figure}[h!]
\includegraphics[width=7.cm]{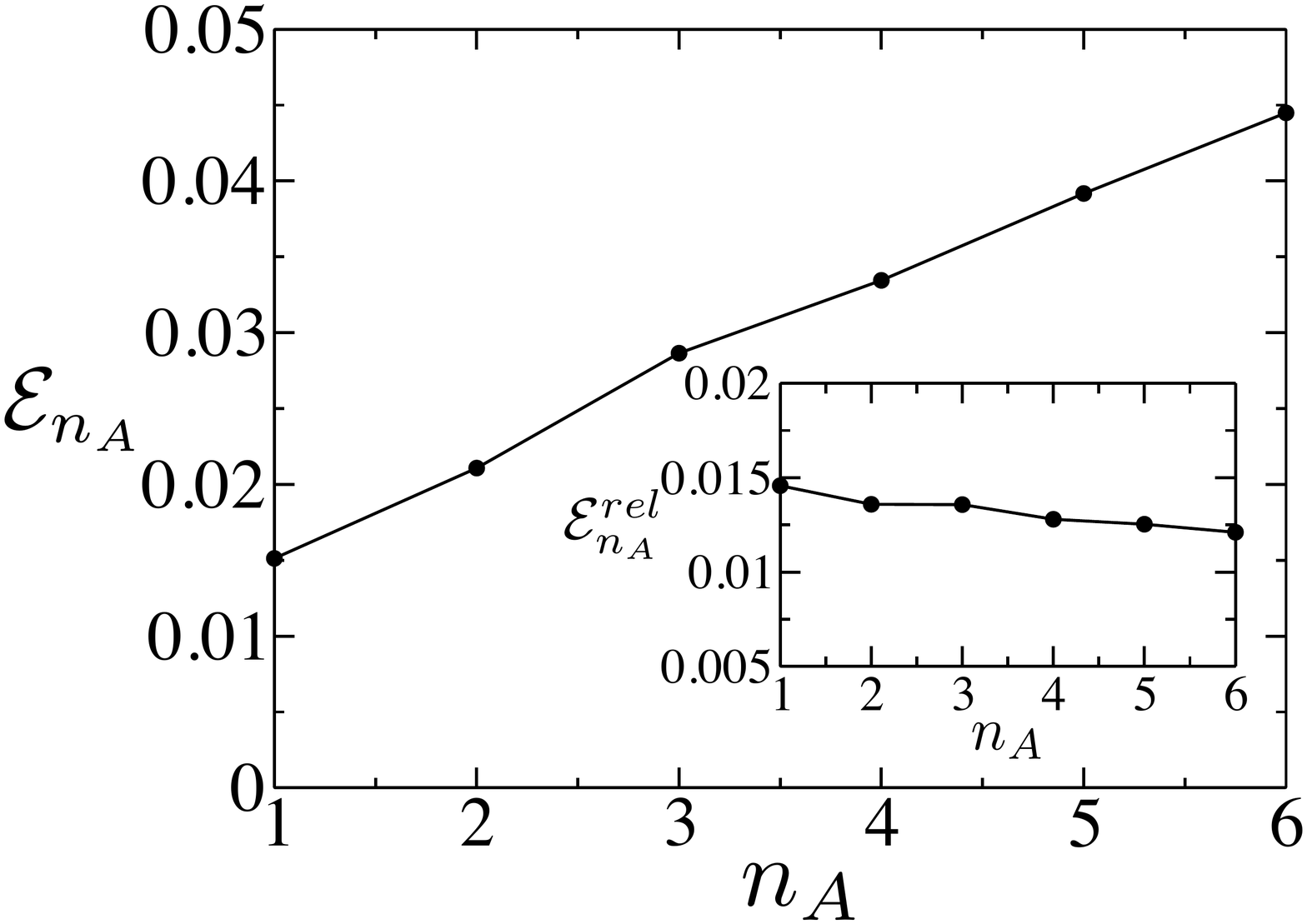}
\caption{Plot of distribution of average error $(\mathcal{E}_{n_A})$      with the number of boundaries between the blocks ($n_A$),  obtained by exploring all possible bipartition of the ground state of the free-fermionic Hamiltonian, defined in Eq. (11) of the main text. In the inset, we also plot the relative error of the entropy $(\mathcal{E}_{n_A}^{rel}$), computed using Eq. (\ref{eq:rel_error}) with the number of boundaries ($n_A$).  Here, $t_{i,i+1}=1$, and  $N=14$.}
\label{fig:error}
\end{figure}

\section{Estimation of error and random sampling }
\label{Appendix_A}
In the first part of this section, we provide a more fine-grained analysis of the error introduced in the computation of the entropies of each bipartition. In particular, we study how the average error is distributed over different geometries of the bipartitions.  Towards this aim, we first group the bipartitions according to the number of boundaries ($n_A$) they share with the rest of the system. The average error ($\mathcal{E}_{n_A}$) is then computed using the following relation
\beq  {\cal E}_{n_A}={1\over Z_A} \sum_{X \in {0, 2^{N-1}}} \left| S_X- \hat S_X \right|,\label{eq:avg_error}
\eeq
where the summation has been taken on the set of bipartitions $X$, which has the same number of boundaries, $n_A$ and $Z_A$ is the cardinality of that set.  Fig. \ref{fig:error} shows the scaling of the error ($\mathcal{E}_{n_A}$) with the number of boundaries $(n_A)$ of the different bipartitions for a free-fermionic model described in Eq. (11) of the main text. From the plot, we can see that the estimated average error is low for bipartitions with less number of boundaries. Now as the bipartitions with more boundaries  yield higher entropy values, in the inset, we also provide the relative error introduced in the computation of the entropies. It is defined as
\beq  {\cal E}_{n_A}^{rel}={1\over Z_A} \sum_{X \in {0, 2^{N-1}}} \frac{\left| S_X- \hat S_X \right|}{S_X}.
\label{eq:rel_error}
\eeq
We note that the relative error is distributed almost uniformly over bipartitions with all possible geometries.

\begin{figure}[h!]
\includegraphics[width=7.5cm]{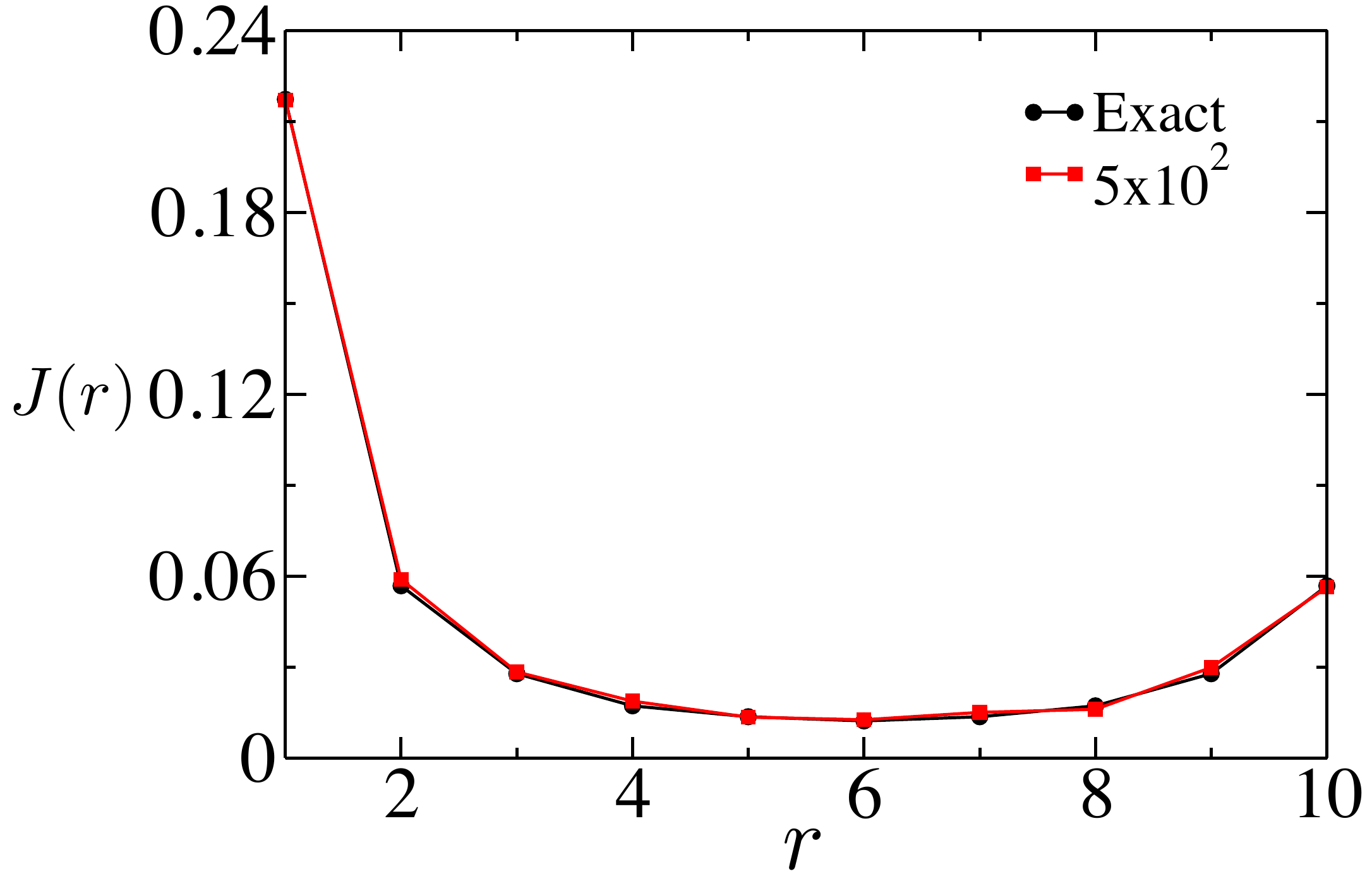}
\caption{Comparison  of decay of entanglement adjacency matrix, $J(r)$ with the site distance $r=|i-j|$ for an $XXZ$ model as given in Eq. (22) of the main text, with  $\Delta =0.50$, considering all possible bipartitions (black curve), and for   sampling of $5 \times 10^2$ number of entropies (red curve). Here, $N=12$.}
\label{fig:random_sampling}
\end{figure}

In the second part of this section,  we discuss the efficiency of the random sampling method that has been introduced in Sec. \ref{sec:numerical} of the main text.  In particular, we compare the behavior of $J(r)$ with the site distance $r=|i-j|$, that obtained in the main text by considering all possible bipartitions of the state, to that obtained using a random sampling of less number of entropies. Fig. \ref{fig:random_sampling} shows one such example where a comparison of $J(r)$ vs $r$ profile, obtained using the random sampling method, has been made to that of the exact method, for  the $XXZ$ model. From the figure, we note that in the cases, random sampling of less number of entropies,  can capture all the features of the entanglement adjacency matrix efficiently.

\section{List of constraints to be fulfilled by an entanglement
  contour}

\label{Appendix_B}

\begin{figure*}[btp]
\includegraphics[width=14.0cm]{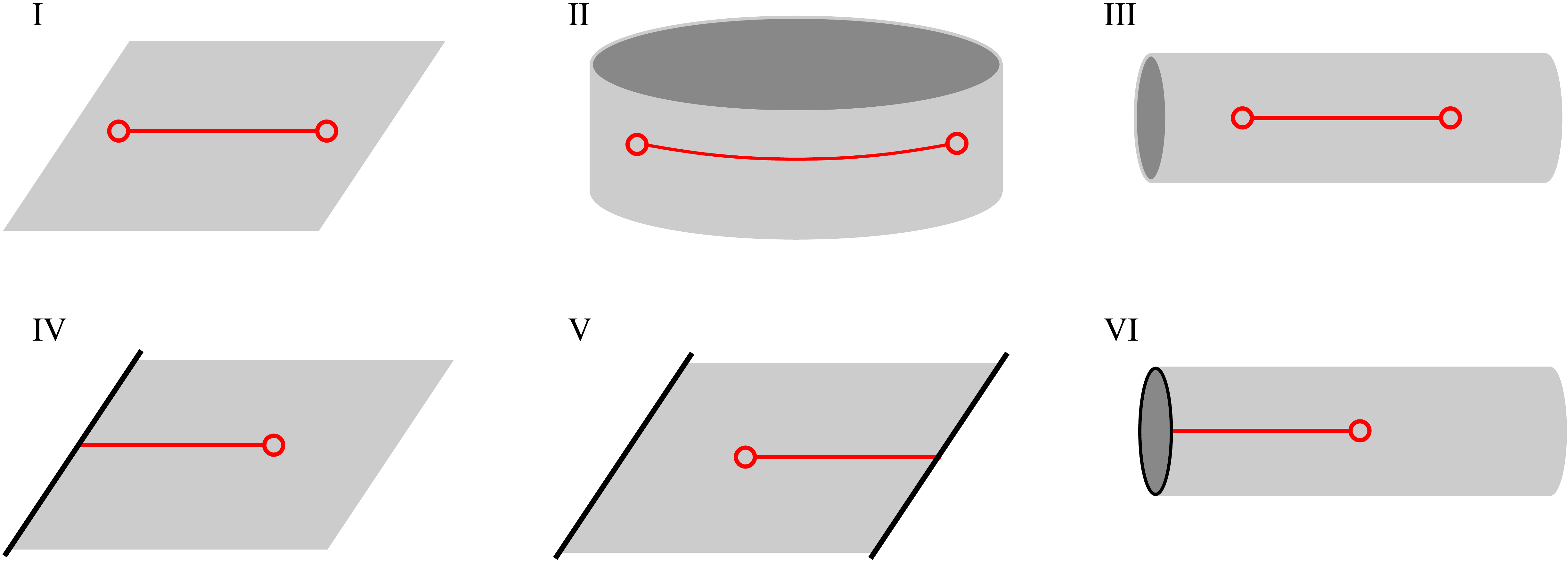}
\caption{Euclidean space-time $\Sigma$ domains of the models. The
  space direction, $x$, runs horizontally and the time direction, $t$,
  runs vertically.  A red segment denotes the interval $A$ and a red
  circle the boundary of a disk of infinitesimal radius $\epsilon$.
  The boundaries of $\Sigma$ are colored in black.  In all these
  cases, the space-time left after the removal of the disk, or discs,
  can be mapped to an annulus $\AN$, of heigh $2 \pi$ and width $W_A$,
  by the conformal transformations given in Table \ref{table1}.  }
\label{figCFT} 
\end{figure*}

\begin{table*}
\begin{tabular}{| c | c  | c | c | c |  c|}
\hline
Type  & Geometry & $A$ & $f(z)$ & $W_A$ & $g(x,y)$  \\
\hline
I & $x,t \in \Rmath$   &$(u,v)$   & $ \log  \frac{z - u}{v - z}$ & $2 \log \frac{v-u}{\epsilon}$  & $\frac{1}{(x-y)^2}$ \\
\hline
II & $x=x+L,  y \in \Rmath$   & $(-R,R)$   & $ \ln \left( \frac{ e^{ 2 \pi i z/L} - e^{ -2 \pi i R/L} }{ e^{ 2 \pi i R/L} - e^{ 2 \pi i z/L} } \right)$ 
& $2 \log \left( \frac{ L}{\pi \epsilon} \sin \frac{ \pi \ell}{L} \right)$  & $\left( \frac{ \pi}{ L} \right)^2  \frac{ 1}{ \sin^2   \frac{  \pi (x-y) }{L}  }$ \\
\hline 
III  & $x \in \Rmath, t = t + \beta$   &$(-R,R)$   &$ \log \left( \frac{ e^{ 2 \pi z/\beta} - e^{ - 2 \pi R/\beta}} {  e^{ 2 \pi R/\beta} - e^{  2 \pi z/\beta}} \right)$ 
& $ 2 \log \left( \frac{ \beta}{ \pi \epsilon} \sinh  \frac{ \pi \ell }{\beta}   \right)$ 
&$\left( \frac{\pi}{\beta} \right)^2   \frac{ 1}{ \sinh^2   \frac{  \pi (x-y) }{\beta}  }$ \\
\hline
IV   & $x > 0, t \in \Rmath$ & $(0, x_0)$ & $\log  \frac{z +x_0}{x_0  - z} $ & $ \log \frac{ 2 x_0}{\epsilon}$ & $ \frac{1}{(x-y)^2} +  \frac{1}{(x+y)^2} $  \\
\hline
V   & $x \in (-L,L), t \in \Rmath$ & $(x_0, L)$ & $\log \Big(\frac{ \sin( \pi ( z - x_0)/(2 L))}{  \cos( \pi ( z + x_0)/(2 L))}\Big) $ 
& $\log \left( \frac{ 4 L}{ \pi \epsilon} \sin \frac{ \pi \ell}{ 2 L} \right) $ 
& $\left( \frac{ \pi}{ 4 L} \right)^2  \Big( \frac{ 1}{ \sin^2 \frac{ \pi (x-y)}{4L} } +  \frac{ 1}{ \sin^2 \frac{ \pi (x+y)}{4L} } \Big) $  \\
\hline
VI   & $x > 0, t = t + \beta$ & $(0,x_0)$ 
& $\log \left( \frac{ e^{ 2 \pi z/\beta} - e^{ - 2 \pi x_0/\beta}} {  e^{ 2 \pi x_0/\beta} - e^{  2 \pi z/\beta}} \right)$ 
& $\log \left( \frac{ \beta}{ \pi \epsilon} \sinh  \frac{ \pi x_0 }{\beta}   \right)$ & $\left( \frac{ \pi}{\beta} \right)^2  
\Big( \frac{ 1}{ \sinh^2 \frac{ \pi(x-y)}{\beta}}  +  \frac{ 1}{ \sinh^2 \frac{ \pi(x+y)}{ \beta} }\Big)$  
\\
\hline
\end{tabular}
\caption{Conformal maps $f(z)$ from the geometries of
  Figs. \ref{figCFT} into the annulus of widths $W_A$ and the
  corresponding functions $g(x,y)$. In the cases $II, III, V$, the
  length of $A$ is denoted by $\ell$.}
\label{table1}
\end{table*}

In this section, we discuss two other important properties to be
satisfied by the contour function \cite{V14} obtained from the
entanglement adjacency matrix, as expressed in Eq. (19) in the main
text.

\begin{enumerate}

\item {\it Invariance under local unitary transformations:} If the
  unitary transformation, $U_B$, concerns a region $B$, and the pure
  state $|\psi\rangle_{AB}$ is related as
  $|\psi^{\prime}\rangle_{AB}=U_B |\psi\rangle_{AB}$, the entanglement
  contour functions $s_A(i)$ derived for both the states
  $|\psi\rangle_{AB}$ and $|\psi^{\prime}\rangle_{AB}$ must be the
  same.

Since the $J_{ij}$ matrix is obtained from the entanglement entropies
of {\em all} blocks, this statement is only true for unitary operators
$U_B$ which act on single sites. In that case, all entropies are
invariant under the action of any local unitary operator and,
therefore, the $J_{ij}$ matrix inherits that invariance.

\item {\it a) Upper bound:} If a subregion $A_1 \subseteq A$ is
  contained in a factor space $V_A=V_{A_1} \otimes V_{A_2}$  then the
  entanglement contour of subregion $A_1$ cannot be larger than the
  entanglement entropy $S_A$, i.e., $s_A(A_1) \leq S_A.$\
  
For a factorized space such as given above, $V_A=V_{A_1} \otimes
V_{A_2}$, the entropy function satisfies the subadditivity condition
$S_A\leq S_{A_1}+S_{A_2}$. This implies $S_{A_1}\leq S_A$ . Again, the
entanglement contour of subregion $A_1$,  $s_A(A_1)=\sum_{i \in A_1, j
  \notin A_1} J_{ij}=S_{A_1}$. Hence the proof. \

Now if we decompose $A_1$ further as $V_{A_1}=V_{A_1^1} \otimes
V_{A_1^2}$, the lower bound can be derived as follows. \

b) {\it Lower bound:} The entanglement contour of subregion $A_1$ is
at least equal to the entanglement entropy of $S_{A_1^1}$.\

This we can show again using the subadditivity condition for the above
factored space, $S_{A_1}\leq S_{A^1_1}+S_{A^2_1}$. Hence, $S_{A_1} \geq
S_{A^1_1}$. Now alternatively, we can write $S_{A_1}=\sum_{i\in A_1 j
  \notin A_1}J_{ij}=s_A(A_1)$. This implies, $s_A(A_1) \geq
S_{A^1_1}$.
\end{enumerate}

\section{Entanglement current in CFT}
\label{Appendix_C}

Let us consider an interval  $A$  of finite length in a larger system. 
The entanglement entropy $S_A$  of the ground state, or the thermal state,  is computed using a path integral 
in  a euclidean space-time $\Sigma$. To regularize the path integral one removes  infinitesimal discs $D_\epsilon$
(parameterized by a complex coordinate $z$) of radius $\epsilon$, centered around the entangling points of $A$.
The resulting space-time $\Sigma \setminus D_\epsilon$, can be mapped, via a conformal transformation
$w = f(z)$,  into an annulus $\AN$ of heigh $2 \pi$ (i.e.  ${\rm Im} \; w = {\rm Im} \; w  + 2 \pi$) and  width  $W_A$ given by \cite{CT16}
\beq 
W_A = \int_{A_\epsilon} dx \, f'(x)  \, , 
\label{cft1}
\eeq
where $A_\epsilon$ denotes  the interval left after the removal of the discs $D_\epsilon$.
The $n^{\rm th}$ R\'enyi entropy  is   given by 
\beq
S_A^{(n)} = \frac{c}{12} \left( 1 + \frac{1}{n} \right) W_A + C_n + o(1) \, , 
\label{cft2}
\eeq
where $c$ is the central charge of the CFT and $C_n$ is a constant that depends on the boundary
entropies and non-universal  data of the model. A contour   is a non-negative  function $s^{(n)}_A(x)$
that describes the contribution of the points of the interval $A$ to the $n^{\rm th}$ R\'enyi entropy \cite{V14}, 
\beq
S_A^{(n)} = \int_{A_\epsilon} dx \; s^{(n)}_A(x), \quad  s^{(n)}_A(x) \geq 0  \, .
\label{cft3}
\eeq
This function is non-unique but  Eqs. (\ref{cft1}) and (\ref{cft2}) suggest the  ansatz \cite{T17,rainbow_bunch4} 
\beq
s_A^{(n)}(x)  = \frac{c}{12} \left( 1 + \frac{1}{n} \right) f'(x) + \frac{C_n}{ \ell}  \, , 
\label{cft4}
\eeq
where $\ell$ is the length of the interval $A$. Our aim is to represent the non-constant term of 
the contour function   (\ref{cft4})  as 
\beq
 s_A^{(n)}(x) - \frac{C_n}{ \ell}    =   \frac{c}{12} \left( 1 + \frac{1}{n} \right) f'(x)= \int_{\bar{A}} dy \;  J^{(n)}(x,y)  \, , 
\label{cft5}
\eeq
where $\bar{A}$ is the complement of $A$.
%
%

The dependence of $J^{(n)}(x,y)$  with respect to  $n$ can be factor out  defining  the function $g(x,y)$,
\beq
J^{(n)}(x,y) = 
 \frac{c}{12} \left( 1 + \frac{1}{n} \right)   g(x,y)  \, , 
\label{cft7}
\eeq
that from  (\ref{cft5}) satisfies 
\beq
 f'(x)  = \int_{\bar{A}} dy \;  g(x,y)  \, . 
\label{cft8}
\eeq
As an example,  let us consider the euclidean space-time with complex coordinate $z = x + i t \in \Cmath$
and the finite interval,  at $t=0$, $A =(u, v) \subset \Rmath$ with $v > u$. 
Removing two discs of radius $\epsilon$ around the entangling points $z = u, v$
one obtains the   annulus $\AN$   by the conformal transformation   \cite{CT16} 
\beq
w = f(z) =  \log  \frac{z - u}{v - z} \, .
\label{cft9}
\eeq 
Choosing the regularized interval as $A_\epsilon = (u+ \epsilon, v - \epsilon)$ (with $0 <\epsilon \ll 1 $),
one finds from (\ref{cft1}) and (\ref{cft9}) 
\beq
W_A = f(v- \epsilon) - f(u+ \epsilon) =  2 \log \frac{ v - u}{\epsilon} \, . 
\label{cft10}
\eeq
and from (\ref{cft2}) the well known result 
\beq
S_A^{(n)} = \frac{c}{6} \left( 1 + \frac{1}{n} \right)  \log \frac{ v - u}{\epsilon}  + C_n + o(1) \, . 
\label{cft11}
\eeq
To find the function $g(x,y)$,  satisfying Eq. (\ref{cft8}),  we 
choose the interval $\bar{A} =( - \infty, u) \cup (v , \infty)$, which does no depend on  $\epsilon$ 
since the interval $A_\epsilon$ already provides a regularization. Eq. (\ref{cft8}) becomes 
\beq
 f'(x)  =  \frac{1}{ x - u} + \frac{1}{ v - x} =   \left( \int_{- \infty}^u + \int_{v}^\infty \right)   dy \;  g(x,y)  \, , 
\label{cft12}
\eeq
and taking a derivative respect to $u$,  or $v$,   gives 
\beq
g(x,y) = \frac{1}{(x-y)^2} \, . 
\label{cft13}
\eeq
that together with (\ref{cft7}) yields (\ref{cft3_text}).  Let us
observe that $J(x,y)$ is the two point correlator of a current ${\bf
  J}(z)$ on the complex plane,

\beq
J(x,y) = \langle {\bf J}(x) \, {\bf J}(y) \rangle_{\rm plane}  \,  . 
\label{cft14}
\eeq
This method to compute $g(x,y)$ can be applied to the geometries
depicted in Fig.  \ref{figCFT} where the regularized space-time
$\Sigma \setminus D_\epsilon$ is conformally equivalent to an annulus
\cite{CT16}.  The results are presented in Table \ref{table1} and
using them one can verify Eq. (28) in the main text that generalizes
(\ref{cft13}).


\subsection*{Two disjoint intervals in the infinite line}

Let us apply   Eqs.  (\ref{cft2_text})  and (\ref{cft3_text}) in the main text to compute the entanglement entropy 
of the ground state for  two disjoint intervals $A = (u_1, v_1) \cup (u_2, v_2)$ (with $u_1 < v_1 < u_2 < v_2$). 
Choosing the regularization 
\barray 
A_\epsilon  & =  &   (u_1 + \epsilon, v_1 - \epsilon) \cup (u_2+ \epsilon , v_2- \epsilon),   \label{68} \\ 
\bar{A}  & =  &   ( - \infty, u_1)  \cup  (v_1, u_2)  \cup ( v_2, \infty)   \, .  \nonumber 
\earray 
one finds (discarding the cutoff $\epsilon$) 
\beq
S_A =  \frac{c}{3} \ln \left( 
\frac{ |u_1 - v_1| |u_2 - v_2 | |u_1 - v_2| |u_2 - v_1|      }{ |u_1 - u_2 | |v_1 - v_2 |}  \right).  \, \,
\label{66}
\eeq
This results agrees with the one first obtained in \cite{CC04}. There
is however a missing additive term in this expression of $S_A$ that
depends on the harmonic ratio $x = (u_1 - v_1)(u_2 - v_2)/(( u_1 -
u_2) (v_1- v_2))$, and the operator content of the CFT, and not only
on the central charge
\cite{CG08,F09,cct-09,cct-11,coser-many,C18}. Extensive analytic and
numerical work has been devoted to this problem but, to our knowledge,
there is no a general analytic formula for $S_A$ (see \cite{C18} for a
summary and an extended list of references). The generalization of
(\ref{66}) to more than two intervals is straightforward.



\begin{thebibliography}{99}
\bibitem{note1} Since both the empty and the complete block should
  present zero entropy, thus they will be left out whenever $s_0
  \neq0$.
\bibitem{note2}
 In the similar vein as that of entanglement contour, we propose an ansatz for the elements of the $J$ matrix as follows.
 \beq
J_{i j} = \left\langle\hspace{-1.5mm}\left\langle \sum_{p=1}^{ {\rm min} (|A|, |\bar{A} | ) } |\Phi_{p,i}^{(A)} \, \Phi_{p,j}^{(\bar{A})} |^2  \;   H(\nu_p)  \
\right\rangle\hspace{-1.5mm}\right\rangle, 
\label{eq:vidal_contour2}
\eeq
%
where $ \Phi_{p,j}^{(\bar{A})}$ is the eigenfunction, with eigenvalue
$\nu_p$, of the correlation matrix \eqref{eq:corr} restricted to the
block $\bar{A}$, and the average is taken over all bipartitions.
Employing (\ref{eq:vidal_contour2}) to find $J_{ij}$ presents some
numerical instabilities which do not arise when using
(\ref{eq:normaleqs}). This equation however provides a physical
interpretation of $J_{ij}$ for free-fermion models.
\end{thebibliography}
\end{document}